\documentclass{iau379}

\usepackage{amsmath}
\usepackage{graphicx}
\usepackage{multirow}

\begin{document}

\lefttitle{Thater et al.}
\righttitle{IAU Symposium 379}

\jnlPage{1}{7}
\jnlDoiYr{2023}
\doival{10.1017/xxxxx}

\aopheadtitle{Proceedings of IAU Symposium 379}
\editors{P. Bonifacio,  M.-R. Cioni, F. Hammer, M. Pawlowski, and S. Taibi, eds.}

\title{Dynamical modelling of ATLAS$^{\rm 3D}$ galaxies}

\author{Thater S. $^1$, Jethwa P.$^1$, Lilley E. J.$^1$, Zocchi A.$^1$, Santucci G.$^{2,3}$ and van de Ven G.$^1$}
\affiliation{$^1$ Department of Astrophysics, University of Vienna, T\"urkenschanzstraße 17, 1180 Vienna\\
$^2$ International Centre for Radio Astronomy Research (ICRAR), M468, University of Western Australia, 35 Stirling Hwy, Crawley, WA 6009, Australia\\
$^3$ ARC Centre of Excellence for All Sky Astrophysics in 3 Dimensions (ASTRO 3D), Australia}

\begin{abstract}
Triaxial dynamical models of massive galaxies observed in the ATLAS$^{\rm 3D}$ project can provide new insights into the complex evolutionary processes that shape galaxies. The ATLAS$^{\rm 3D}$ survey is ideal as the sample comprises a good mix of fast and slow rotators with vastly different mass assembly histories. We present a detailed dynamical study with our triaxial modelling code DYNAMITE, which models galaxies as a superposition of their stellar orbits. The models allow us to constrain the intrinsic shape of the stellar component, the distributions of the visible and invisible matter and the orbit distribution in these nearby early-type galaxies and to relate it with different evolutionary scenarios. Triaxial modelling is essential for these galaxies to understand their complex kinematical features. 
\end{abstract}


\maketitle

\section{Introduction}

Early-type galaxies (ETGs) are distinguished into
two classes based on their apparent angular momentum: fast rotators and slow rotators \citep[e.g.][]{Emsellem2007, Cappellari2007,Emsellem2011}. While fast rotators are nearly axisymmetric and often have an oblate shape, slow rotators are weakly triaxial (but not far from isotropic). These two classes likely depict two different channels of galaxy formation. Slow rotators are thought to have assembled in the centres of massive halos and after an intense star formation period at high redshift have evolved from gas-poor major mergers. In contrast, fast rotators have likely formed out of star-forming discs and their evolution is dominated by gas accretion, bulge growth and quenching \citep{Cappellari2016}. In this study we investigate the evolutionary histories of ATLAS$^{\rm 3D}$ galaxies by searching for imprints in their dynamically inferred intrinsic shapes and stellar orbit distributions.


\section{The ATLAS$^{\rm 3D}$ sample}

ATLAS$^{\rm 3D}$ \citep{Cappellari2011} is a multiwavelength survey that includes 260 ETGs with stellar masses $M_*>6 \times 10^9$ M$_{\odot}$ within the local volume (42\:Mpc). The sample was deduced from a parent sample which was carefully selected to be statistically representative of the nearby galaxy population. About 25\% of the ATLAS$^{\rm 3D}$ ETGs were classified as elliptical galaxies and 75\% as lenticular galaxies. The survey was carried out with the SAURON integral field unit on the William Herschel Telescope. From these observations, detailed 2-dimensional stellar kinematic maps of mean velocity, velocity dispersion, and $h_3$ and $h_4$ Gauss-Hermite polynomials with high S/N are available, which are ideal for our dynamical study. The kinematics \citep[e.g.][]{Krajnovic2011,Emsellem2011}, star formation history \citep[e.g.][]{McDermid2015} and environment \citep[e.g.][]{Cappellari2011,Serra2012} of the ATLAS$^{\rm 3D}$ galaxy sample have been extensively studied in the past decade, allowing us to put our dynamical results in the context of galaxy assembly histories.

\section{First dynamical results with DYNAMITE}

\begin{figure}[t]
\centering
\includegraphics[width=\textwidth]{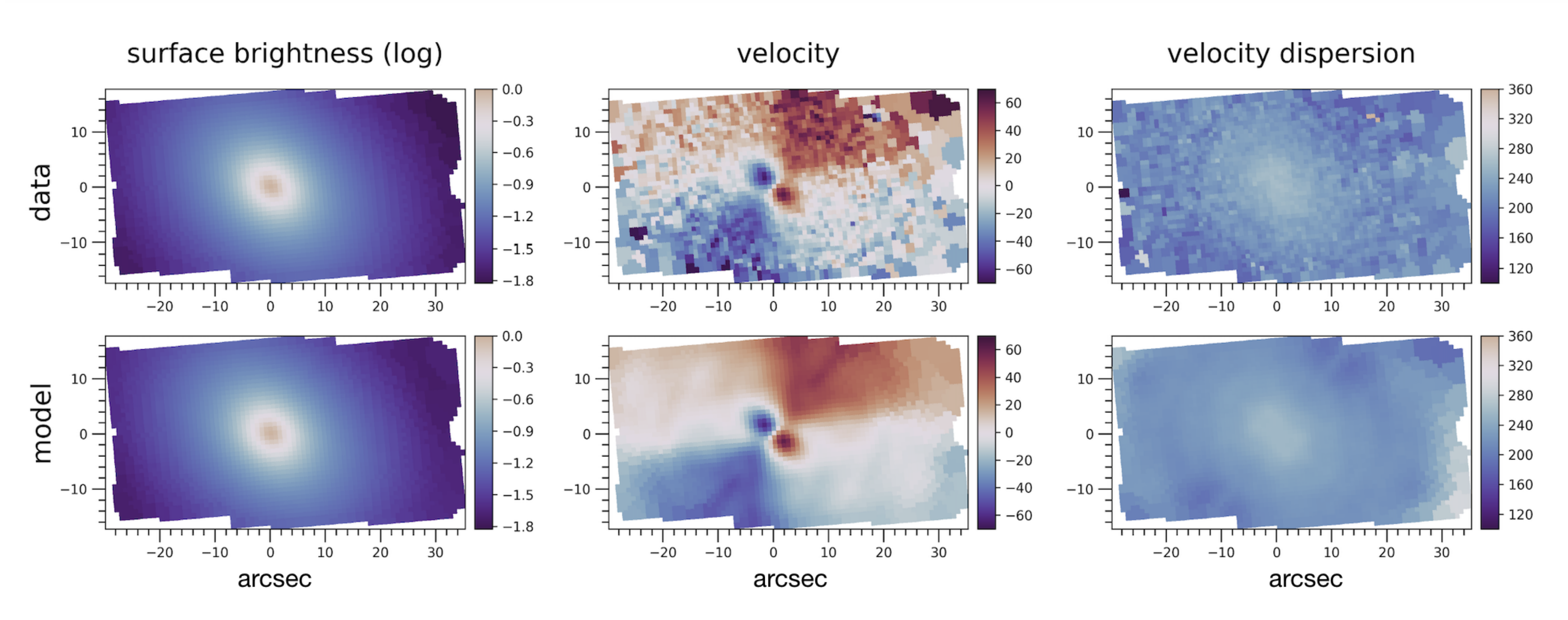} 
\caption{DYNAMITE makes it possible to dynamically model complex triaxial
features such as kinematically decoupled components. We show here the surface brightness, mean velocity and velocity dispersion of the SAURON data (top) and of the Schwarzschild model (bottom) of the ATLAS$^{\rm 3D}$ galaxy NGC 4365.}
\label{ff:kin}
\end{figure}

We obtained the data products from the ATLAS$^{\rm 3D}$ webpage\footnote{\url{http://www-astro.physics.ox.ac.uk/atlas3d/}} and for the first time constructed triaxial Schwarzschild orbit-superposition models for these galaxies using DYNAMITE\footnote{\url{https://dynamics.univie.ac.at/dynamite_docs/}} ( \citealt{vandenbosch2008, Jethwa2020, Thater2022}). DYNAMITE allows us to recover the enclosed mass, the intrinsic shape of the stellar component, the dark matter fraction and the orbit distribution of the modelled galaxies. So far, we have finished modelling 63 galaxies, about a quarter of the full ATLAS$^{\rm 3D}$ sample. Figure~\ref{ff:kin} shows a comparison between the observed SAURON kinematics and the best-fit DYNAMITE model of NGC 4365. NGC 4365 harbours a kinematically decoupled core (KDC) in its centre that is remarkably well recovered by our models. In the ATLAS$^{\rm 3D}$ survey, many galaxies are exhibiting complex kinematic features which require triaxial Schwarzschild modelling, and not only regularly rotating galaxies (that can be well modelled assuming axisymmetry).
 
We modelled the ATLAS$^{\rm 3D}$ galaxies with the same modelling setup as described in \cite{Santucci2022} for SAMI galaxies. Our gravitational potential consists of a stellar component, a central black hole, and dark matter parametrised as a spherical halo with a Navarro-Frenk-White \citep[NFW;][]{Navarro1996} radial profile. In total, we consider five free parameters: a constant stellar mass-to-light ratio $M_*/L$, intrinsic stellar axis-length ratios $p$ (intermediate-to-long) and $q$ (short-to-long), the stellar projected-to-intrinsic scale-length ratio $u$ and the dark matter fraction $f_{\rm DM} = M_{200}/M_*$ within the virial radius $r_{200}$. The dark matter concentration $c$ was fixed using the $M_{200} - c$ relation by \cite{Dutton2014} to get a better handle on the degeneracy between $M_*/L$ and dark matter. The ATLAS$^{\rm 3D}$ kinematics do not have the spatial resolution to constrain the central black hole mass ($M_{\rm BH}$), therefore we fixed black hole masses using the empirical $M_{\rm BH} - \sigma_{\rm e}$ relation by \citet{vandenbosch2016}. From our modelling approach, we obtain dynamical masses that are consistent with those obtained by \citet{Cappellari2013} and \citet{Poci2017} using Jeans modelling.

\begin{figure}
\centering
\includegraphics[width=0.85\textwidth]{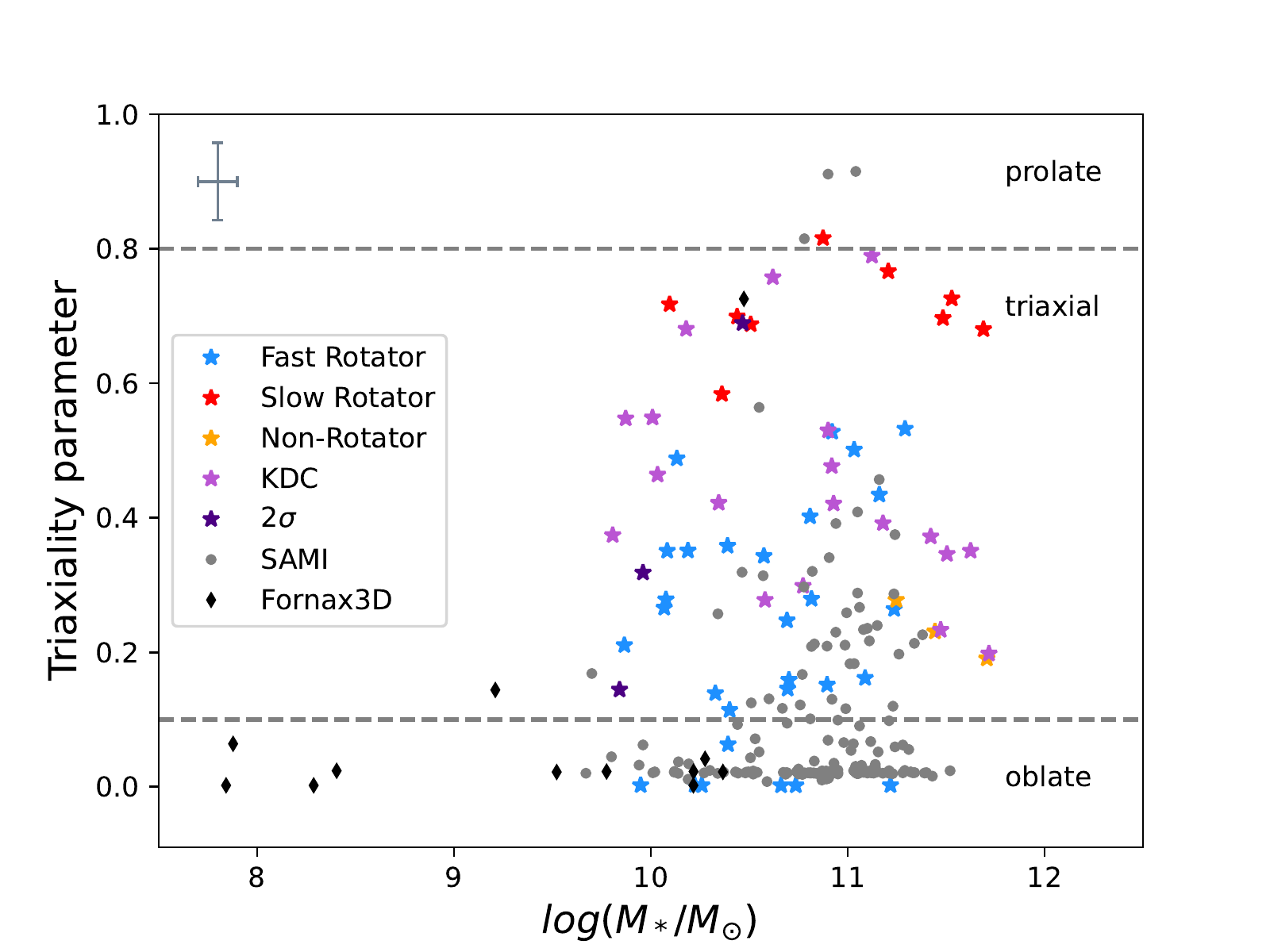} 
\caption{Triaxiality parameter $T_{\rm Re}$ as a function of stellar mass. We colour-code the ATLAS$^{\rm 3D}$ galaxies based on their kinematic classification from \cite{Krajnovic2011}:  regular fast-rotating galaxies are represented here in blue, featureless slow-rotators in red, non-rotators in orange, KDCs in light purple, and 2$\sigma$ galaxies in dark purple. Typical uncertainties at $1\sigma$ confidence level are shown in the top left corner of the plot. For comparison, we also show two other samples of galaxies that were modelled with DYNAMITE, from the SAMI \citep{Santucci2022} and Fornax3D \citep{Ding2023} surveys, represented here with grey dots and black diamonds, respectively.}
\label{ff:shape}
\end{figure}

In this study, we aim at constraining the intrinsic shape of the stellar component of massive ETGs. The triaxiality parameter at one effective radius ($R_{\rm e}$) is defined as \mbox{$T_{\rm Re}=(1-p_{\rm Re}^2)/(1-q_{\rm Re}^2)$}. Galaxies with $T_{\rm Re} = 0$ are classified as oblate, galaxies with \mbox{$T_{\rm Re} = 1$} as prolate and galaxies with values of $T_{\rm Re}$ between 0.1 and 0.8 as triaxial. In Figure~\ref{ff:shape}, we show the triaxiality parameter as a function of stellar mass; the ATLAS$^{\rm 3D}$ galaxies we modelled are shown as stars in the plot, with different colours indicating the kinematic classification by \citet[][]{Krajnovic2011}. While most of our modelled galaxies have triaxial intrinsic shapes, we find trends that depend on their kinematic classification. Most of the modelled galaxies are regularly fast-rotating galaxies (blue) and have a mildly triaxial shape ($\overline{T_{\rm Re}} = 0.25$). However, the scatter is quite strong and some galaxies have oblate or strongly triaxial shapes. Featureless slow-rotators ($\overline{T_{\rm Re}} = 0.71$, red) on the other hand are strongly triaxial or even prolate (which is also seen in the kinematics). Interestingly, the modelled non-rotating galaxies (orange; only 3 galaxies) have a mildly triaxial shape ($\overline{T_{\rm Re}} = 0.23$) similar to fast-rotators. 
The ATLAS$^{\rm 3D}$ survey contains about 20 galaxies with KDCs. For these galaxies we find on average a strongly triaxial intrinsic shape ($\overline{T_{\rm Re}} = 0.45$) which is consistent with these galaxies having mainly formed through mergers. Surprisingly, we also find strong triaxiality for the three $2\sigma$ galaxies (galaxies exhibiting two peaks in the velocity dispersion). As this feature is typically found in galaxies with a pair of extended, counter-rotating discs, we would have expected these galaxies to be more oblate. The triaxial shape might be a remnant signature of their merger history. More galaxies and a deeper analysis are essential to confirm our first results. In contrast to our modelled ATLAS$^{\rm 3D}$ galaxies, the majority of modelled galaxies in the SAMI  \citep{Santucci2022} and Fornax3D \citep{Ding2023} surveys have oblate shapes, as shown in Fig.~\ref{ff:shape} (with grey dots and black diamonds, respectively).

\begin{figure}
\centering
\includegraphics[width=0.60\textwidth]{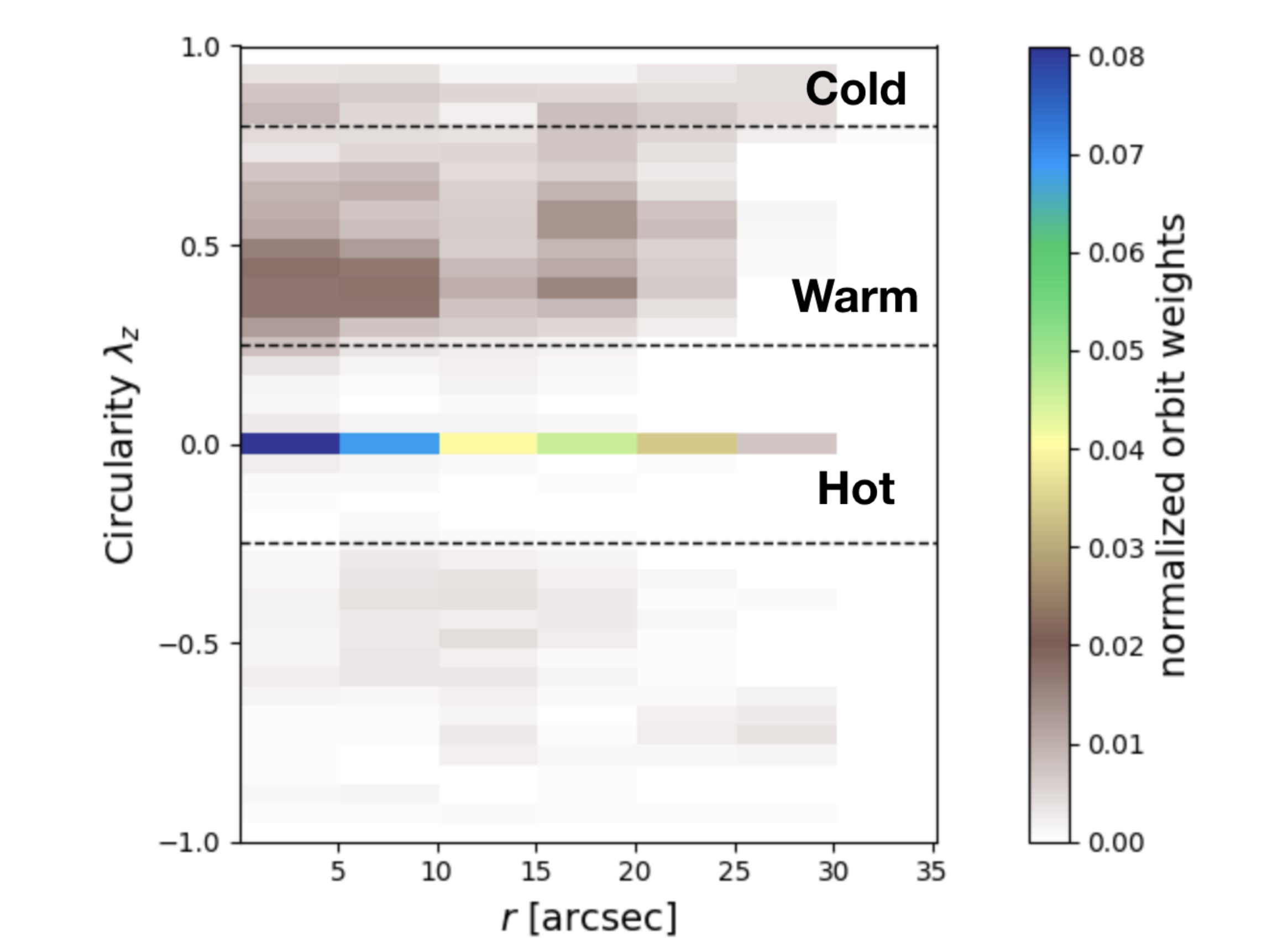} 
\caption{Stellar orbit distribution of the fast-rotating galaxy NGC 821. Stellar orbits are characterized by two main properties: the time-averaged radius $r$, representing the size of each orbit, and the circularity $\lambda_z$, which compares the angular momentum of the orbit with the one of a circular orbit. The density of the orbits in phase space is normalized to unity within the data coverage ($\sim 30$ arcsec) and indicated by the different colours. The three horizontal black dashed lines indicate $\lambda_z$ = 0.8, $\lambda_z$ = 0.25 and $\lambda_z$ = -0.25, dividing the orbits into four regions (cold, warm, hot and counter-rotating orbits).}
\label{ff:orbit}
\end{figure}

DYNAMITE models also allow a detailed look at the distribution of the stellar orbits of the galaxies. In Fig~\ref{ff:orbit}, we show the orbit distribution of NGC 821, one of our fast-rotating ETGs. As expected for bulge-dominated galaxies, we see a high fraction of hot orbits with a circularity $\lambda_{\rm z} \sim 0$ (consistent with a pressure-supported bulge). However, there is also a large fraction of warm and cold orbits, meaning that the ETG exhibits significant net angular momentum. We will compare the orbit distribution for galaxies belonging to the various kinematic classes and search for signatures of accretion as in \cite{Zhu2020, Zhu2022}.

In the future, we will extend our dynamical study to the remaining galaxies of the ATLAS$^{\rm 3D}$ survey and search for correlations between the intrinsic shapes/orbit distributions and kinematic properties, environment and star formation histories, in order to better understand the formation and evolutionary processes of early-type galaxies.
\\\\
This research was supported by the European Union's Horizon 2020 research and innovation programme under grant agreement NO 724857 (Consolidator Grand ArcheoDyn).


\begin{thebibliography}{}
\bibitem[Cappellari et al.(2007)]{Cappellari2007} Cappellari, M., Emsellem, E., Bacon, R., et al.\ 2007, MNRAS, 379, 418
\bibitem[Cappellari et al.(2011)]{Cappellari2011} Cappellari, M., Emsellem, E., Krajnovi{\'c}, D., et al.\ 2011, MNRAS, 413, 813
\bibitem[Cappellari et al.(2013)]{Cappellari2013} Cappellari, M., Scott, N., Alatalo, K., et al.\ 2013, MNRAS, 432, 1709
\bibitem[Cappellari(2016)]{Cappellari2016} Cappellari, M.\ 2016, ARAA, 54, 597
\bibitem[Ding et al.(2023)]{Ding2023} Ding, Y., Zhu, L., van de Ven, G., et al.\ 2023, A\&A, 672, A84
\bibitem[Dutton \& Macci{\`o}(2014)]{Dutton2014} Dutton, A.~A. \& Macci{\`o}, A.~V.\ 2014, MNRAS, 441, 3359
\bibitem[Emsellem et al.(2007)]{Emsellem2007} Emsellem, E., Cappellari, M., Krajnovi{\'c}, D., et al.\ 2007, MNRAS, 379, 401
\bibitem[Emsellem et al.(2011)]{Emsellem2011} Emsellem, E., Cappellari, M., Krajnovi{\'c}, D., et al.\ 2011, MNRAS, 414, 888
\bibitem[\protect\citeauthoryear{Jethwa et al.}{2020}]{Jethwa2020} Jethwa P., Thater S., Maindl T., van de Ven G., 2020, ascl.soft. ascl:2011.007
\bibitem[Krajnovi{\'c} et al.(2011)]{Krajnovic2011} Krajnovi{\'c}, D., Emsellem, E., Cappellari, M., et al.\ 2011, MNRAS, 414, 2923
\bibitem[McDermid et al.(2015)]{McDermid2015} McDermid, R.~M., Alatalo, K., Blitz, L., et al.\ 2015, MNRAS, 448, 3484
\bibitem[Navarro et al.(1996)]{Navarro1996} Navarro, J.~F., Frenk, C.~S., \& White, S.~D.~M.\ 1996, ApJ, 462, 563
\bibitem[Poci et al.(2017)]{Poci2017} Poci, A., Cappellari, M., \& McDermid, R.~M.\ 2017, MNRAS, 467, 1397
\bibitem[Santucci et al.(2022)]{Santucci2022} Santucci, G., Brough, S., van de Sande, J., et al.\ 2022, ApJ, 930, 153
\bibitem[Serra et al.(2012)]{Serra2012} Serra, P., Oosterloo, T., Morganti, R., et al.\ 2012, MNRAS, 422, 1835
\bibitem[\protect\citeauthoryear{Thater et al.}{2022}]{Thater2022} Thater S., Jethwa P., Tahmasebzadeh B., et al., 2022, A\&A, 667, A51
\bibitem[van den Bosch et al.(2008)]{vandenbosch2008} van den Bosch, R.~C.~E., van de Ven, G., Verolme, E.~K., et al.\ 2008, MNRAS, 385, 647
\bibitem[van den Bosch(2016)]{vandenbosch2016} van den Bosch, R.~C.~E.\ 2016, ApJ, 831, 134
\bibitem[\protect\citeauthoryear{Zhu et al.}{2020}]{Zhu2020} Zhu L., van de Ven G., Leaman R., et al., 2020, MNRAS, 496, 1579
\bibitem[\protect\citeauthoryear{Zhu et al.}{2022}]{Zhu2022} Zhu L., van de Ven G., Leaman R., et al., 2022, A\&A, 664, A115

\end{thebibliography}
\end{document}